\input{aipcheck}

\documentclass[
    ,final            
  ]
  {aipproc}

\layoutstyle{8x11single}

\newcommand{\la}{\langle}
\newcommand{\ra}{\rangle}

\newcommand{\beq}{\begin{eqnarray}}
\newcommand{\eeq}{\end{eqnarray}}

\newcommand{\bfl}{\begin{flushleft}}
\newcommand{\efl}{\end{flushleft}}
\newcommand{\boldsymbol}{\bf}

\begin{document}

\title{
${D_{s1}^{*}(2710)}$ and $D_{sJ}^{*}(2860)$ in the $\widetilde{U}(12) \times O(3,1)$-scheme
}

\classification{14.40.Lb,13.25.Ft,12.39.Ki}
\keywords      {$D_{s1}^{*}(2710)$, $D_{sJ}^{*}(2860)$, Strong Decays, Covariant Oscillator Quark Model, the $\widetilde{U}(12) \times O(3,1)$-scheme}

\author{Tomohito Maeda}{
  address={Department of Engineering Science, 
Junior College Funabashi Campus, Nihon University,\\ Funabashi 274-8501, Japan}
}

\author{Kenji Yamada}{
  address={Department of Engineering Science, 
Junior College Funabashi Campus, Nihon University,\\ Funabashi 274-8501, Japan}
}

\author{Masuho Oda}{
  address={School of Science and Engineering, 
Kokushikan University,\\ Tokyo 154-8515, Japan}
}

\author{Shin Ishida\footnote{Senior Research Fellow}~~}{
  address={Research Institute of Science and Technology, 
College of Science and Technology, Nihon University,\\ Tokyo 101-8308, Japan}
}

\begin{abstract}
In order to classify the charmed-strange mesons,   
including puzzling $D_{s0}^{*}(2317)$ and 
$D_{s1}(2460)$, and recently observed  
${D_{s1}^{*}(2710)}$ and $D_{sJ}^{*}(2860)$, we employ 
the $\widetilde{U}(12) \times O(3,1)$ level-classification 
scheme of hadrons proposed and developed by us in recent years. 
The scheme has a new degree of freedom, $SU(2)_{\rho}$, which 
leads to a number of extra states out of conventional 
$SU(6)$-scheme, named {\it chiralons}. 
Applying this novel classification scheme, we investigate 
the strong decays of $S$- and $P$-wave $c\bar{s}$ mesons 
with one pseudoscalar emission by using the covariant 
oscillator quark model. 
As a result it is shown that $D_{s0}^{*}(2317)$ and 
$D_{s1}(2460)$ are described as 
${}^{1}S_{0}^{\chi}$ and ${}^{3}S_{1}^{\chi}$ 
chiralons, forming the $SU(2)_{\rho}$ doublet with the 
ground-state $D_{s}^{}$ and 
$D_{s}^{*}$, respectively.  
Furthermore, the observed decay properties of 
$D_{s1}^{*}(2710)$ is consistently explained as 
the vector chiralon ${}^{1}P_{1}^{\chi}$. 
On the other hand, it is also found that the controversial 
narrow state, $D_{sJ}^{*}(2860)$, does not fit as predicted  
properties of our $P$-wave vector chiralon. 
\end{abstract}

\maketitle

\section{Introduction}
In recent years we have proposed the $\widetilde{U}(12)\times O(3,1)$ 
level-classification scheme{~\cite{Ishida:2000-2002}} of hadrons, 
which corresponds to a covariant 
extension of the non-relativistic $SU(6)\times O(3)$-scheme. 
In the new scheme, wave functions (WF) of composite hadrons are generally given 
by irreducible tensors of the $\widetilde{U}(12)\times O(3,1)$.  
It is to be noted that, at the rest frame of hadrons, the representations 
of the $\widetilde{U}(12)$ 
reduce to those of the $U(12)$. 
Hence the hadronic states can be classified by representations of the $U(12)$.  
The $U(12)$ includes a new degree 
of freedom $SU(2)_{\rho}$, called {\it $\rho$-spin}, in addition to 
conventional $SU(6)$ ($\supset SU(2)_{\sigma} \times SU(3)_{F}$), 
as $U(12) \supset SU(6) \times SU(2)_{\rho}$. 
An important feature of the scheme is that we include, apart from the $O(3,1)$ part, the `$\rho$-spin down' 
($\rho_{3}=-1$) component, which is treated as a fundamental 
building block{\footnote{Physical meaning of $\rho_{3}= -1$ component is discussed 
in the Ref.~{\cite{Yamada:2010}}.} to construct the WF, effectively realizing only inside hadrons. 
Accordingly, we expect to exist a number of extra states 
out of the $SU(6)$-framework, which is called chiralons.
{\footnote{More strictly, chiralons are defined as the 
states which includes at least one $\rho_{3}=-1$ 
component in the spin WF.}}

To check the validity of our new classification scheme, 
we investigate systematically the strong decays of $c\bar{s}$ 
mesons with one pseudoscalr emission by using the covariant 
oscillator quark model (COQM). Through the observed mass and 
results of decay study we present possible new assignments 
for observed charmed-strange mesons from the view point of 
the $\widetilde{U}(12) 
\times O(3,1)$-scheme.  
\section{Framework of the covariant oscillator quark model}
We briefly recapitulate a framework 
of the COQM{\footnote{
The COQM have been successfully applied 
to various (static and non-static) problem of hadrons for many years. 
As relatively recent application, for example, see 
Ref.~{\cite{Oda et al.:2000}} and references therein. }}
relevant to the present application. 
The WF of a composite $s\bar{c}$ meson system is described by a  
bilocal field ${\Phi}(X, x)_{\alpha}{}^{\beta}$, (and its Pauli conjugate defined by 
$\bar{\Phi}(X, x)_{\alpha}{}^{\beta}=(\gamma_{4})_{\alpha^{}}{}^{\alpha^{'}}
(\Phi^{\dagger})_{\alpha^{'}}{}^{\beta^{'}} (\gamma_{4})_{\beta^{'}}{}^{\beta}$), where 
$\alpha$ and $\beta$ represent Dirac spinor indices  
of $s$- and $\bar{c}$-quark, $X$ and $x$ denote the center of mass (CM) and relative 
coordinate four vectors, respectively. 

We start from the Klein-Gordon type equation, 
\beq 
\left(\frac{\partial^2}{\partial X_{\mu}^2}-{\cal M}(x)^2 \right) 
{\Phi}(X, x)_{\alpha}{}^{\beta}=0.
\eeq
The squared-mass operator{
\footnote{
By imposing the definite metric-type subsidiary 
condition to freeze a redundant relative-time degree of freedom, 
we get the eigen-values of ${\cal M}(x)^2$ as 
$M_{n}^2= n\Omega + M_{0}^2$, where $\Omega=d\sqrt{\frac{K}{\mu}}$ and 
$n=L+2N$ ($N$ and $L$ 
being the radial and orbital quantum numbers respectively), 
leading linear rising Regge trajectories. 
}} (in the pure-confining harmonic oscillator (HO) potential limit) 
is given by 
\beq
\label{Eq:2}
{\cal M}(x)^2 = d \left[-\frac{1}{2\mu}\frac{\partial^2}
{\partial x^{2}_{\mu}}+\frac{K}{2} x_{\mu}^2 \right] \ \ \ 
 \left( d\equiv {2(m_{s}+m_{\bar{c}})}, \ \ \mu \equiv 
\frac{m_{s}m_{\bar{c}}}{m_{s}+m_{\bar{c}}}\right).
\eeq
We can define the plane wave expansion concerning the CM motion 
\beq
\Phi(X, x)_{\alpha}{}^{\beta}= \int \frac{d^{3} 
{\bf P}}{\sqrt{(2\pi)^{3} 2 P_{0}}} 
( e^{+i PX} \Phi(x, P)^{(+)}_{\alpha}{}^{\beta} 
+ e^{-i PX} \Phi(x, P)^{(-)}_{\alpha}{}^{\beta} ), 
\eeq
with respect to each level ($M_{n}=\sqrt{-P_{\mu}^2}$)  
determined by the squared-mass eigen-equation. 
In the above the positive / negative frequency parts $\Phi(x, P)^{(+/-)}$ 
denotes the internal WF 
of relevant ($s\bar{c}$) mesons with a definite HO mass. 
The complete set of (boosted) $LS$-coupling basis, generaly represented by 
\beq
\Phi(x,P)_{\alpha}^{(\pm) }{}^{\beta}\sim  f_{\mu_{1}\mu_{2}\cdots}(v, x)  \otimes
\left(W(v)_{\alpha}^{(\pm)}{}^{\beta}\right)_{\mu_{1}\mu_{2}\cdots}, 
\eeq
is used to expand the internal WF, 
where $f(v, x)$ indicates the space-time part, while $W(v)_{\alpha}^{(\pm)}{}^{\beta}$ 
does the spin part. Here $v_{\mu}=P_{\mu} / M $ is four velocity of meson. 
The concrete expressions of former part $f(v,x)$, 
being the eigen functions of ${{\cal M}^2}$, are given by 
\beq
f_{G}(v, x)=\frac{\beta}{\pi}
\exp\left(-\frac{\beta}{2}\left(x_{\mu}^{2}+2(v_{\mu}x_{\mu})^2\right)\right)
\stackrel{{\bf v=0}}{\rightarrow} \frac{\beta}{\pi} \exp \left(-\frac{\beta}{2}\left(\boldsymbol{x}^2+x_{0}^{2}\right)\right) \ \ (
\beta=\sqrt{\mu K})
\eeq
for $S$-wave ground states and  
\beq
f_{\nu}(v, x)&=& a^{\dagger}_{\nu}f_{G}(x,v) = 
\frac{1}{\sqrt{2\beta}}(\beta x_{\nu}-\frac{\partial}{\partial x_{\nu}})f_{G}(x,v) 
=\sqrt{2\beta}\left(x_{\nu}+v_{\nu}(x_{\rho}v_{\rho})\right)f_{G}(x,v)
\eeq
for $P$-wave excited states, respectively. On the other hand, the later part 
$W(v)^{(\pm)}$ consists of the direct product of respective 
Dirac spinor bases, which simulates the transformation properties of relevant 
constituent quarks, as  
\beq
W_{r, r^{'}}^{(+)}(v)_{\alpha}{}^{\beta}
=u^{(s)}_{r}(v)_{\alpha}~\bar{v}_{r^{'}}^{(\bar{c})}(v)^{\beta}, \ \  \ \ \
W_{r^{}, r^{'}}^{(-)}(v)_{\alpha}{}^{\beta}
=v^{(s)}_{r^{}}(v)^{\alpha}~\bar{u}_{r^{'}}^{(\bar{c})}(v)^{\beta}.
\eeq
Here the index $r $ represents the eigenvalue of 
$\rho_{3} ${\footnote{
For the anti-quark spinor $v_{r}(v)$, it should be understood as 
$\bar{\rho}_{3}=-\rho^{t}_{3}$.}} 
in the rest frame (${\bf v}={\bf 0}$). 
It should be noted that these spinors do not correspond to 
constituent quarks themselves. In fact, these contain only four velocity of hadron, 
hence the small component vanishes at the hadron rest frame. 
\section{$S$- and $P$-Wave $s\bar{c}$ Mesons in the $\tilde{U}(12) \times O(3,1)$-scheme}
In this work we make the following assumptions 
for spin WF; only $\rho_{3}=+1$ 
is allowed for c-quark, while both $\rho_{3}=\pm 1$ can be realized for 
(rather) light s-quark.{\footnote
{In the Ref. {\cite{Yamada:2010}}, $\rho_{3}=-1$ component for $c$-quark is 
taken into account. 
}}
Resultant WF of $S$-wave states are given by 
\beq
\label{Eq:8}
\Phi(x,P)^{(+)}_{\alpha}{}^{\beta}
&=&f_{G}(v, x)~ 
\left[W^{(+)}_{+,+}(v)|_{{(S=0)}}+W^{(+)}_{+,+}(v)|_{{(S=1)}}
+W^{(+)}_{-,+}(v)|_{{(S=0)}}+W^{(+)}_{-,+}(v)|_{{(S=1)}}
\right]_{\alpha}{}^{\beta}\nonumber\\
&=&f_{G}(v, x)~ 
\frac{1}{2\sqrt{2}} \left[ \left(i\gamma_{5} \bar{D}_{s}(P)+i\gamma_{\mu} 
\bar{D}^{*}_{s\mu}(P)+
\bar{D}_{s 0}^{\chi}(P)+i\gamma_{5}\gamma_{\mu} \bar{D}^{\chi}_{s 1\mu}(P) \right)
(1+\frac{iP{\cdot}\gamma}{M_{0}})
\right]_{\alpha}{}^{\beta},  
\eeq
where $\{ \bar{D}_{s}, \bar{D}^{*}_{s\mu}, \bar{D}_{s 0}^{\chi}, \bar{D}^{\chi}_{s 1\mu} \}$ 
represent local $s\bar{c}$ meson 
fields with $J^{P} =\{ 0^{-}, 1^{-}, 0^{+}, 1^{+} \}$, respectively. 
A superscript $\chi$ implies {\it chiralon}, $s$-quark being $\rho_{3}=-1$. 
It should be noted that, in the relevant case, chiralons always have 
their `partners' with opposite parity, same spin $J$, forming the $\rho$-spin doublet. 
{\footnote{Clearly degeneracy of mass (in the HO limit) for the $SU(2)_{\rho}$ 
doublets is badly broken. 
Thus the $SU(2)_{\rho}$ should be considered, 
in contrast to the $SU(2)_{\sigma}$-symmetry, 
just to offer a tool which gives a new perspective on classifying hadronic states. 
}}
Similarly, WF of $P$-wave states are given by 
\beq
\label{Eq:9}
\Phi(x,P)^{(+)}_{\alpha}{}^{\beta}&=&f_{\nu}(v, x)~ 
\left[\left(W^{(+)}_{+,+}(v)|_{{(S=0)}}+W^{(+)}_{+,+}(v)|_{{(S=1)}}
+W^{(+)}_{-,+}(v)|_{{(S=0)}}+W^{(+)}_{-,+}(v)|_{{(S=1)}}\right)_{\nu}
\right]_{\alpha}{}^{\beta}\nonumber\\
&=&
\sqrt{2\beta}x_{\nu}f_{G}(v, x)~ 
\frac{1}{2\sqrt{2}} \left[ \left(i\gamma_{5} \bar{D}^{}_{s 1 \nu}(P)+
i\gamma_{\mu} \bar{D}^{*}_{s J \mu\nu}(P)+
\bar{D}_{s 1 \nu}^{~\chi}(P)+i\gamma_{5}\gamma_{\mu} \bar{D}^{* \chi}_{s J \mu\nu}(P) \right)
(1+\frac{iP{\cdot}\gamma}{M_{1}})
\right]_{\alpha}{}^{\beta}, \nonumber\\
\eeq
where the local fields $\{ \bar{D}_{s 1 \nu}, \bar{D}^{*}_{s J\mu\nu}, 
\bar{D}_{s 1 \nu}^{\chi}, \bar{D}^{\chi}_{s J\mu\nu} \}$ 
correspond to $J^{P} =\{ 1^{+}, \{J= 0, 1, 2\}^{+}, 1^{-}, \{J= 0, 1, 2 \}^{-} \}$ states, 
respectively{
\footnote{
All Lorentz indices of local fields satisfy the subsidiary conditions; 
$v_{\mu}D_{s\mu}^{*}=v_{\mu}D_{s1\mu}^{\chi}=
v_{\mu}{D}^{}_{s 1 \mu}=v_{\mu}{D}^{*}_{s J \mu\nu}=v_{\mu}{D}_{s 1\mu}^{~\chi}
=v_{\mu}{D}^{* \chi}_{s J \mu\nu}=0$, ~${D}^{*}_{s 0,~2 \mu\nu}={D}^{*}_{s 0,~2 \nu\mu}, ~ 
{D}^{*}_{s 1 \mu\nu}=-{D}^{*}_{s 1 \nu\mu}$, and ${D}^{*}_{s 2 \mu\mu}=0$. 
}}. 
\section{Pionic / Kaonic decays}
Next we explain a procedure for calculating the pionic / kaonic 
decays of $D_{s}$ mesons, applying the COQM. 
It can be considered that decays proceed through a single quark 
transition via emission of a local pion / kaon. 
We introduce the decay interactions as follows:     
\beq
S_{\rm int}=\int d^4 x_{1} \int d^{4}x_{2}~
\la\bar{\Phi}^{(-)}(x_{1}, x_{2})V(x_{1} )\Phi^{(+)}(x_{1}, x_{2}) \ra,
\eeq
where $x_{1}$ and $x_{2}$ denote the space-time coordinates of $s$- and $\bar{c}$-quarks 
related to CM and relative 
coordinates as  
$X_{\mu}=(m_{s}x^{}_{1\mu}+m_{c}x^{}_{2\mu})/(m_{s}+m_{{c}})$,  
$x_{\mu}=x^{}_{1\mu}-x^{}_{2\mu}$, and $\la \cdots \ra$ means taking trace concerning 
flavor and Dirac indices. 
Two types of vertex factors,  
$V(x_{1})=V_{ND}(x_{1})+V_{D}(x_{1})$, 
denoting non-derivative and derivative couplings{
\footnote{
In the non-relativistic limit, the first term contributes only 
the transitions between chiralons and non-chiralons, accompanied by $\rho_{3}$-change. 
On the other hand, the second term contributes only transitions among non-chiralons 
or chiralons themselves, which gives the well known ${\bf \sigma}\cdot ({\bf q}-\frac{q_{0}}{m_{q}}{\bf p}_{q})$ vertex. 
}},  
are 
\beq
&&\la\bar{\Phi}^{(-)}(x_{1},x_{2})V_{ND}(x_{1})\Phi^{(+)}(x_{1},x_{2})\ra
= dg_{ND}\la\bar{\Phi}^{(-)}(x_{1},x_{2})
\left(i\gamma_{5}\phi_{\rm ps}(x_{1})\right)\Phi^{(+)}(x_{1},x_{2})\ra ,\\
&&\la \bar{\Phi}^{(-)}(x_{1},x_{2}) V_{D}(x_{1}) \Phi^{(+)}(x_{1},x_{2})\ra 
=\frac{d g_{D}}{2m_{s}}
\la\bar{\Phi}^{(-)}(x_{1},x_{2})
\left(\overrightarrow{\partial}_{x_{1\mu}} \phi_{\rm ps}(x_{1})\right)
\left(\gamma_{5}\sigma_{\mu\nu}(\overrightarrow{\partial}_{x_{1}}-\overleftarrow{\partial}_{x_{1}})\right)\Phi^{(+)}(x_{1},x_{2})\ra. \nonumber\\
\eeq
Rewriting the above with CM and relative coordinates 
by 
\beq
\Phi^{(+)}(x_{1}, x_{2})\sim \Phi^{(+)}(x,P)e^{+iP\cdot X},  \ \ 
\bar{\Phi}^{(-)}(x_{1}, x_{2}) \sim \bar{\Phi}^{(-)}(x,P)e^{-iP^{'}\cdot X}
\eeq 
and 
\beq
\phi_{{\rm ps}}(x_{1}) \sim  
\phi_{{\rm ps}}(q) e^{-iq\cdot x_{1}}=
\left(\begin{array}{ccc}\frac{\pi^{0}}{\sqrt{2}}+\frac{\eta^{(8)}}{\sqrt{6}} & \pi^{+} & K^{+} \\\pi^{-} & -\frac{\pi^{0}}{\sqrt{2}}+\frac{\eta^{(8)}}{\sqrt{6}} & K^{0} \\K^{-} & \bar{K}^{0} & -\sqrt{\frac{2}{3}}\eta^{(8)}\\ \end{array}\right)
e^{-iq\cdot (\frac{2m_{c}}{d}x+X)}  \ \ \ (q_{\mu}=P_{\mu}-P^{'}_{\mu}), 
\eeq
we obtain a formula to calculate 
the decay amplitudes as  
\beq
T&=&dg_{ND}\int d^{4} x \la \bar{\Phi}^{(-)}(P^{'},x)i\gamma_{5} \phi_{{\rm ps}}(q) 
\Phi^{(+)}(P, x) 
\ra e^{-i\frac{2m_{c}}{d}q \cdot x}\\
&&+g_{D} \int d^{4} x \la \bar{\Phi}^{(-)}(P^{'},x)\gamma_{5}q_{\mu}\sigma_{\mu\nu}
\left( P_{\nu}+P^{'}_{\nu}-\frac{d}{2m_{s}}i(\overrightarrow{\partial}_{x,\nu}-\overleftarrow{\partial}_{x,\nu}) \right) \phi_{{\rm ps}}(q) \Phi^{(+)}(P, x) \ra
e^{-i\frac{2m_{c}}{d}q \cdot x}. 
\eeq
\begin{table}
 \label{table1}
 \begin{tabular}{cc|cc|cccc} 
\hline 
$~n~$  &  $~L~$  &  $P$  &   $V$   & $S^{\chi}$   & $A^{\chi}$ \\ 
\hline
  &  & {$1^{1}S_{0}$}  & {$1^{3}S_{1}$} &  {$1^{1}S^{\chi}_{0}$}&  {$1^{3}S_{1}^{\chi}$} \\
$0$ &  $0$  &  $0^{-}$  &  $1^{-}$   &    $0^{+}$  &  $1^{+}$ \\  
\cline{3-6}
 &  &   \multicolumn{2}{c|}{$({2.11}\ \mathrm{GeV})$} &     \multicolumn{2}{c}{$({2.46}\ \mathrm{GeV})$}  \\
	         &          &  $D_{s}(1968)$   &  $D^{*}_{s}(2112)$   & $D^{*}_{s0}(2317)$    &  $D_{s1}(2460)$\\ \hline
	         &          & {$1^{1}P_{1}$}&  {$1^{3}P_{J=0,1,2}$} &  {$1^{1}P_{1}^{\chi}$}&
{$1^{3}P_{J=0,1,2}^{\chi}$} \\
	$1$   &  $1$  &  $1^{+}$    &  $\{0,1,2\}^{+}$   &  $1^{-}$    &  $\{0,1,2\}^{-}$ \\  \cline{3-6}  
	&          & \multicolumn{2}{c|}{$({2.57} \ \mathrm{GeV})$}  &
\multicolumn{2}{c}{$(2.87 \ \mathrm{GeV})$}
\\	         &        & $D^{}_{s1}(2536)$  &  
\underline{$D^{*}_{s0}(\sim 2573)$}, \underline{ $D^{}_{s1}(\sim 2573)$}, $D_{s2}(2573)$  &  $D_{s1}^{*}(2710)$  &  
\underline{$D^{}_{s0}(\sim 2866)$}, $D^{*}_{sJ}(2860)$?, \underline{$D^{}_{s2}(\sim 2866)$ }\\ 
\hline
\end{tabular}
\caption{Possible assignments of $S$- and $P$- wave $D_{s}$ mesons in the $\widetilde{U}(12)_{SF}\times O(3,1)$-scheme}  
\end{table}
\begin{table}
\begin{tabular}{l|c|c|c|c|c} \hline
Mesons & ${}^{2S+1}L_{J};{j_{q}^P}$ & $(\rho_{3}(c),\bar{\rho}_{3}(\bar{s}))$ 
& decay channel&$\Gamma^{\rm theor.}$&$\Gamma^{\rm exp.}$\\
\hline
$D_{s}$ & ${}^{1}S_{0};{\frac{1}{2}}^{-}$ & $(+,+)$ & - & - &- \\
$D_{s}^{*}$ & ${}^{3}S_{1};{\frac{1}{2}}^{-}$ & $(+,+)$ & $D_{s}~\pi^{0}$ & 0.0020 keV&$<$110 keV\\
$D_{s0}^{*\chi}(2317)$ & ${}^{1}S_{0}^{\chi};{\frac{1}{2}}^{+}$ & $(+,-)$ & $Ds~\pi^{0}$ & 9.2 keV&$<$3.8 MeV\\
$D_{s1}^{\chi}(2460)$ & ${}^{3}S_{1}^{\chi};{\frac{1}{2}}^{+}$ & $(+,-)$ & $Ds^{*}~\pi^{0}$ & 8.2 keV& $<$ 3.5 MeV \\
\underline{$D_{s0}(\sim 2573)^{*}$} & ${}^{3}P_{0};{\frac{1}{2}}^{+}$ & $(+,+)$ & $D~K$ & $\sim$ 184 MeV &\\
\underline{$D_{s1}(\sim 2573)$} & ${}^{}P_{1};{\frac{1}{2}}^{+}$ & $(+,+)$ & $D^{*}~K$ &$\sim$ 184 MeV &\\
$D_{s1}(2536)$ & ${}^{}P_{1};{\frac{3}{2}}^{+}$ & $(+,+)$ & $D^{*}~K$ & 0.22 MeV & $<$ 2.3 MeV\\
$D_{s2}^{*}(2573)$ & ${}^{3}P_{2};{\frac{3}{2}}^{+}$ & $(+,+)$ & $D~K+D^{*}~K$ &18.9+0.96=19.9 MeV &20$\pm$5 MeV \\
$D_{s1}^{\chi}(2710)$ & ${}^{1}P_{1}^{\chi} $ & $(+,-)$ & $D^{}~K+D^{*}K$ & 63 + 43 =106 MeV & $149^{+46}_{-59} $ MeV\\
\underline{$D_{sJ}^{*\chi}(\sim 2860)$} & ${}^{3}P_{1}^{\chi}$ & $(+,-)$ & $D~K+D^{*}~K$ &177+60=237 MeV &$48 \pm 9 $ MeV \\
\hline
\end{tabular}
\caption{
Results on pionic / kaonic transition widths (in MeV) 
}
\label{table2}
\end{table}
\section{Assignments and Numerical Results}
New charmed strange meson $D_{sJ}^{*}(2860)$ 
was first observed by the BaBar collaboration{\cite{BaBar:2006}} 
in the $DK$ channel of $e^{+}e^{-}$ inclusive 
measurement with $M=2856.6 \pm 1.5 \pm 5.0{\rm MeV}$ 
and $\Gamma=48 \pm 7 \pm 10 {\rm MeV}$. 
Shortly after, a $J^{P}=1^{-}$ 
state $D_{s1}^{*}(2710)$ was reported by the 
Belle collaboration{\cite{Belle:2006-2007}} with $M=2708\pm 9^{+11}_{-10}{\rm MeV}$ 
and $\Gamma=108 \pm 23^{+36}_{-31} {\rm MeV}$ 
in the $DK$ invariant mass distribution of $B$-decay. 
In the latest report from the 
BaBar collaboration{\cite{BaBar:2009}}, the $D_{sJ}^{*}(2860)$ and $D_{s1}^{*}(2710)$ 
were seen in both $DK$ and $D^{*}K$ decay modes with the ratios 
of branching fraction
\begin{eqnarray}
\frac{BR(D_{S1}^{*}(2710)\to D^{*}K)}{BR(D_{S1}^{*}(2710)\to D^{}K)}
=0.91\pm 0.13 \pm 0.12 , \ \ 
\frac{BR(D_{SJ}^{*}(2860)\to D^{*}K)}{BR(D_{SJ}^{*}(2860)\to D^{}K)}
=1.10 \pm 0.15 \pm 0.19 .
\end{eqnarray}
To understand the nature of these newly observed states, many 
theoretical efforts have been done, 
\footnote{For example, the preceding work{\cite{Matsuki:2007,Close et al.:2007}} 
predicts $D_{sJ}^{*}(2860)$ and $D_{s1}^{*}(2710)$ as $2P$ and $2S$ (or $1D$) states 
of conventional $c\bar{s}$ states.} 
but it is still a subject of controversy. 

Now we discuss a classification based on 
the $\widetilde{U}(12) \times O(3,1)$-scheme, shown in Table~{\ref{table1}}. 
Subsequent results for pionic / kaonic transition widths 
in comparison with experiments are also given in Table~{\ref{table2}}. 
In both tables, some predicted, but experimentally missing states are underlined. 
\footnote{Though the decay studies we have used following parameters:\\  
(i)~coupling constants; $g_{D}={g_{A}}/{\sqrt{2}f_{\pi~(K)}}$, 
$g_{A} = 0.55$,  $dg_{ND}=8.0 $ GeV. ($f_{\pi~(K)}=94~(114)$ MeV)~~
(ii)~HO ($SU(2)_{\rho}$-symmetric) mass; $M_{0}=2.26$GeV, $M_{1}=2.69$ GeV. ~~
(iii)~Regge slope inverse; $\Omega= 2.160$  GeV$^{2}$,  
determined from $M(D_{s2}^{*}(2573))^2 - M(D_{s}^{*}(2112))^2$.~~
(iv)~qaurk mass ratio; $m_{c}/m_{s}$=1.55/0.45. ~~
(v)~$\eta^{(8)}-\pi_{0}$ mixing angle; $ \sin^2 \theta$ = 
$0.65 \times 10^{-4}${\cite{Cho and Wise:1994}}.}
In consideration of unexpectedly lower mass, $D_{s0}^{*}(2317)$ and $D_{s1}(2460)$ 
are plausible candidates for our $S$-wave chiralons. 
Thus we assign them to $D_{s0}^{\chi}$ and $D_{s1}^{\chi}$ chiralons  
in Eq.~(\ref{Eq:8}), respectively. A issue raised by relevant assignments is whether 
there are additional (conventional) $c\bar{s}$ P-wave states, 
$D_{s1}$ and $D_{s1}^{*}$~{\footnote{
More properly, mixed states 
$|D_{s1}^{j=1/2}\ra = \sqrt{2/3}|D_{s1}^{*}\ra-\sqrt{1/3}|D^{}_{s1}\ra$ and  
$|D_{s1}^{j=3/2}\ra = \sqrt{1/3}|D_{s1}^{*}\ra+\sqrt{2/3}|D^{}_{s1}\ra$ 
are realized in the heavy quark limit. 
}} in Eq.~({\ref{Eq:9}}). The mass of 
those $P$-wave $1^{+}$ non-chiralons are expected to be about $\sim 2.57$ GeV 
from that of typical $D_{s2}^{*}(2573)$, being much 
heavier than $S$-wave chiralons. 
The results of strong decays, assuming the existences of $D^{*}_{s0}(\sim 2.57)$ and 
$D_{s1}^{j=1/2}(\sim 2.57)$ non-chiralons in Table~{\ref{table2}} indicate that 
it is not contradict with present experiment, owing to predicted large widths. 
On the other hand, the experimental known state $D_{s1}(2536)$ is naturally explained as 
$D_{s1}^{j=3/2}$ non-chiralons, being mixing partner of $D_{s1}^{j=1/2}$. 
Next we make a rough estimate the mass of the $P$-wave chiralons, by using global 
HO mass relation,
$M_{1}^2 = \Omega + M_{0}^2$, 
derived from Eq.~({\ref{Eq:2}}).   
As a result, we predict two vector chiralons, $D_{s1}^{\chi}$ and $D_{s1}^{*\chi}$, 
with the mass about $2.7 \sim 2.9$ GeV. These states are possible candidate of 
recently reported $D_{s1}^{*}(2710)$and $D_{sJ}^{*}(2860)$. We calculate the strong 
decays to check this possibility, and found that $D_{s1}^{*}(2710)$ meson 
is consistently explained as 
the vector chiralon ${}^{1}P_{1}^{\chi}$. 
On the other hand, $D_{sJ}^{*}(2860)$, does not fit as predicted  
properties of our $P$-wave vector chiralon. 
\vspace{-0.6em}
\section{Concluding Remarks}
In conclusion, the $D_{s0}^{*}(2317)$, 
$D_{s1}(2460)$, and $D_{s1}^{*}(2710)$ 
are good candidates for $c\bar{s}$ chiralons through 
their observed masses and decay properties. 
The existence of  
$P$-wave non-chiralon $D^{*}_{s0}$ and $D_{s1}$ with broad width ($\sim 180$ MeV) and 
higher mass ($\sim 2.57$ GeV) appeared in the $DK$ and $D^{*}K$ spectrum, respectively, 
and that of $J^{P}=\{0,1,2\}^{-}$ $P$-wave 
chiralons with the mass $2.7 \sim 2.9$ GeV should be checked 
in future experiment. 
\vspace{-1.5em}
\begin{theacknowledgments}
This work was supported in part by Nihon University Research 
Grant for 2008. 
\end{theacknowledgments}


%
\bibliographystyle{aipproc}   

\vspace{-1em}
\begin{thebibliography}{99}
\bibitem{Ishida:2000-2002} 
S. Ishida, M. Ishida and T. Maeda, \emph{Prog. Theor. Phys.}{\textbf{104}} (2000) 785; 
S. Ishida and M. Ishida, \emph{Phys. Lett.} {\textbf{B539}} (2002) 249. 
\bibitem{Yamada:2010}
K. Yamada, in these proceedings. 
\bibitem{Oda et al.:2000}
M. Oda, K. Nishimura, M. Ishida, and S. Ishida, 
\emph{Prog. Theor. Phys.}{\textbf{133}} (2000) 1213.
\bibitem{PDG:2008} C.~Amsler, et al., (Particle Data Group), 
\emph{Phys. Lett.} \textbf{B667} (2008) 1. 
\bibitem{Matsuki:2007}
T. Matsuki, T. Morii, and K. Sudoh, \emph{Eur. Phys. J.} {\textbf{A31}} (2007) 701.
\bibitem{Close et al.:2007}
F.E. Close, C.E. Thomasa, O. Lakhinab and E. S. Swanson, 
\emph{Phys. Lett.} {\textbf{B647}} (2007) 159. 
\bibitem{BaBar:2006} B.~Aubert et al., BABAR Collaboration, \emph{Phys. Rev. Lett.} 
{\textbf{97}} (2006) 222001.  
\bibitem{BaBar:2009}  B.~Aubert, et al., BaBar Collaboration, \emph{Phys. Rev. } 
{\textbf{D80}} (2009) 092003. 
\bibitem{Belle:2006-2007} K.~Abe, et al., Belle Collaboration, hep-ex/0608031; 
J.~Brodzicka et al., Belle Collaboration, \emph{Phys. Rev. Lett.}\textbf{100}(2008) 
092001. 
\bibitem{Cho and Wise:1994} P.~L.~Cho and M.~B.~Wise, \emph{Phys. Rev. }{\textbf{D49}}(1994) 6228. 
\end{thebibliography}

\end{document}